\documentclass[twocolumn]{aastex7}

\usepackage{comment}
\usepackage{graphicx}
\usepackage{hyperref}
\usepackage{afterpage}
\begin{document}

\title{Not All Sub-Neptune Exoplanets Have Magma Oceans}

\author[0009-0000-1065-0654]{Bodie Breza}
\affiliation{Department of Astronomy, University of Maryland, College Park, MD 20742, USA}
\email[show]{bgbreza@umd.edu}

\author[0000-0001-8236-5553]{Matthew C. Nixon}
\altaffiliation{51 Pegasi b Fellow}
\affiliation{Department of Astronomy, University of Maryland, College Park, MD 20742, USA}
\affiliation{School of Earth and Space Exploration, Arizona State University, Tempe, AZ 85287, USA}
\email[show]{matthewnixon@asu.edu}

\author[0000-0002-1337-9051]{Eliza M.-R. Kempton}
\affiliation{Department of Astronomy, University of Maryland, College Park, MD 20742, USA}
\affiliation{Department of Astronomy \& Astrophysics, University of Chicago, Chicago, IL 60637, USA}
\email{}

\begin{abstract}

The evolution and structure of sub-Neptunes may be strongly influenced by interactions between the outer gaseous envelope of the planet and a surface magma ocean. However, given the wide variety of permissible interior structures of these planets, it is unclear whether conditions at the envelope-mantle boundary will always permit a molten silicate layer, or whether some sub-Neptunes might instead host a solid silicate surface. In this work, we use internal structure modeling to perform an extensive exploration of surface conditions within the sub-Neptune population across a range of bulk and atmospheric parameters. We find that a significant portion of the population may lack present-day magma oceans. In particular, planets with a high atmospheric mean molecular weight and large envelope mass fraction are likely to instead have a solid silicate surface, since the pressure at the envelope-mantle boundary is high enough that the silicates will be in solid postperovskite phase. This result is particularly relevant given recent inferences of high-mean molecular weight atmospheres from JWST observations of several sub-Neptunes. We apply this approach to a number of sub-Neptunes with existing or upcoming JWST observations, and find that in almost all cases, a range of solutions exist which do not possess a present-day magma ocean. Our analysis provides critical context for interpreting sub-Neptunes and their atmospheres.

\end{abstract}


\section{Introduction} \label{sec:intro}

Modern exoplanet surveys have revealed an abundance of close-in planets with sizes ranging from 1-4 $R_{\oplus}$ \citep{Batalha_2013}. This population exhibits a bimodal radius distribution, with a first peak at $\sim$1.3 $R_\oplus$, the second at 
$\sim$2.4 $R_\oplus$, and a minimum at $\sim$1.75 $R_\oplus$ \citep{Fulton_2017}. Planets with radii $<$ 1.75 $R_\oplus$ are often referred to as super-Earths, while those with radii $>$ 1.75 $R_\oplus$ are called sub-Neptunes. The bulk densities of sub-Neptunes indicate that they must possess some kind of volatile component, which could take the form of, e.g., a H/He envelope making up a few percent of the planet’s mass \citep{Lopez_2014,Rogers_2015,Chen_2016}, or a water-rich envelope of up to $\sim50\%$ of the planet’s mass \citep{Leger_2004,Zeng_2019,Luque_2022}. However, mass and radius alone are not enough to disentangle these possibilities \citep[e.g.,][]{Rogers_2011,Rogers_2023}. In order to understand the composition of these objects, we require atmospheric characterization, both through observational constraints on the upper atmosphere via spectroscopy \citep[e.g.,][]{Kreidberg_2014,Kempton_2023,Benneke_2024,Davenport_2025}, and through detailed modeling of the key formation and evolutionary processes that shape these atmospheres throughout their lifetimes \citep[e.g.,][]{Owen_2013,Kite_2019,Schlichting_2022}.

One of the key processes that shapes the evolution of sub-Neptunes arises from the fact that their gaseous envelopes are expected to be in contact with a long-lived magma ocean \citep{Kite_2019}. Interactions between the magma and the envelope have been invoked to explain the steep decline in the frequency of sub-Neptunes with radii above 3 $R_{\oplus}$ \citep{Fulton_2017}. This occurs because magma oceans will limit growth due to the solubility of hydrogen in magma at high pressures \citep{Kite_2019}. Furthermore, the interaction between a magma ocean and a H$_2$-rich envelope can affect atmospheric composition and structure \citep[e.g.,][]{Schlichting_2022, Misener_2023, Seo_2024}, potentially leading to observable atmospheric signatures \citep[e.g.,][]{Rigby_2024,Shorttle_2024,Werlen_2025}.

Although magma oceans may be crucial in affecting the atmospheric composition and structure of sub-Neptunes, it is currently unclear whether the silicate layer of a given sub-Neptune is expected to be molten at all times. The thermodynamic conditions at the envelope-mantle boundary will be governed by numerous parameters, including the planet mass, temperature profile and atmospheric composition. Although it has been noted in some cases that certain sub-Neptunes may not currently host magma oceans \citep{Rigby_2024}, there is very little work exploring the parameter space for which magma oceans are expected to persist or solidify. Additionally, molecular dynamics simulations have shown that for sub-Neptunes, hydrogen and water may be distributed in the core and mantle as well as the gaseous envelope \citep{Stixrude2005,Luo2024}. However, in order to determine whether the thermodynamic conditions of these simulations are relevant to sub-Neptune interiors, we must first evaluate the range of possible temperatures and pressures that could occur at the envelope-mantle boundary for planets with different bulk and atmospheric properties.

JWST has revealed a wide range of atmospheric compositions within the sub-Neptune population \citep[e.g.][]{Madhusudhan_2023, Benneke_2024, Piaulet-Ghorayeb_2024,Davenport_2025}. JWST/MIRI observations of sub-Neptune GJ~1214~b \citep{Kempton_2023} found a metal rich upper atmosphere, with \cite{Gao_2023} finding a $\gtrsim$ 300x solar metallicity, implying a mean molecular weight (MMW) of $\gtrsim$ 9.8 g/mol. Additional observations using JWST/NIRSpec also point to a high-MMW atmosphere \citep{Schlawin2024,Ohno2025}. Internal structure modeling informed by the atmospheric constraints from \citet{Kempton_2023} and \citet{Gao_2023} suggests a high envelope mass fraction ($x_{\rm env}$), resulting in high pressures at the envelope-mantle boundary that could push the rocky surface into a solid phase \citep{Nixon_2024}. GJ~1214~b is not the only sub-Neptune observed to have a high-MMW atmosphere, with JWST observations suggesting a MMW $>5\,$g/mol for TOI-270~d \citep{Benneke_2024}, GJ~9827~d \citep{Piaulet-Ghorayeb_2024} and TOI-836~c \citep{Wallack2024}. These findings motivate a wider parameter exploration to determine the prevalence of magma oceans on sub-Neptunes with a range of bulk and atmospheric properties. Due to the important role which magma oceans play in both atmospheric composition and planetary structure, it is vital that we explore the possibility of their absence in a subset of sub-Neptunes. 

In this work, we carry out an important first step towards determining the extent to which magma oceans may be expected to persist in sub-Neptunes. We generate a large grid of steady-state interior structure models and determine the thermodynamic conditions at the envelope-mantle boundary for each. We determine the key planetary properties that influence the likelihood of a magma ocean's existence, and consider the implications of our findings for a number of sub-Neptunes which have recently been observed or are set to be observed with JWST.

\section{Methods} \label{sec:methods}

\subsection{Internal Structure Model}
We use the same internal structure model as in \cite{Nixon_2024}, which is an updated version of the model initially presented in \cite{Nixon_Madhusudhan_2021}, titled Structural Model of Internal Layers of Exoplanets (\texttt{SMILE}\footnote{\texttt{github.com/mcnixon/smile}}). This model solves the mass continuity and hydrostatic equilibrium equations using numerical integration and assuming spherical symmetry. Planets consist of an iron core, silicate mantle, and a mixed H/He/H$_2$O envelope. We solve for the planetary radius $R_p$ for an input mass $M_p$, mass fractions $x_i$ = $M_i$/$M_p$ of each component, and envelope MMW. The mass fractions of H$_2$O and H/He are calculated from the MMW \citep[see equation 1,][]{Nixon_2024}. Throughout this work, we assume an Earth-like iron-to-silicate ratio (1/3 iron, 2/3 silicates by mass).

\texttt{SMILE} uses an isothermal equation of state (EOS) for iron, adopted from \cite{Seager_2007}. Thermal expansion is negligible for iron at high pressures \citep{Howe_2014}. The silicate EOS prescription is temperature-dependent, following the phase diagram shown in \cite{Huang_2022}, and consists of three phases: bridgmanite \citep{Oganov_2004}, postperovskite \citep{Sakai_2016}, and liquid MgSiO$_3$ \citep{Wolf_2018}. The silicate phase diagram can be seen in Figure \ref{fig:GJ 1214b}.

A temperature-dependent EOS is also used for the outer gaseous envelope, as the temperature profile of the envelope can significantly affect the radii of sub-Neptunes \citep{Nixon_Madhusudhan_2021}. The EOS for H$_2$O is adopted from \cite{Nixon_Madhusudhan_2021} and was constructed from a range of sources, covering the full pressure-temperature ($P$--$T$) space which is relevant to sub-Neptunes \citep{Salpeter_1967,Fei_1993,Wagner_2002,Feistel_2006,Seager_2007,French_2009,Klotz_2017,Journaux_2020}. The H/He EOS is adopted from \cite{Chabrier_2019} and assumes a helium mass fraction \textit{Y} = 0.275. Our implementation of the \texttt{SMILE} model assumes a fully mixed H/He/H$_2$O envelope, as predicted by simulations of sub-Neptunes with equilibrium temperatures down to $\sim$300~K \citep{Gupta2025}. An ideal mixing approximation is used to construct the mixed EOS, which has been shown to be appropriate for these components \citep{Soubiran_2015}.

The photospheric temperature ($T_0$) and pressure ($P_0$) are also required input parameters. While we allow $T_0$ to vary as part of our grid, we fix $P_0$ to a value of 100 Pa, or $10^{-3}$ bar. This is a typical photospheric pressure for an sub-Neptune atmosphere in transit geometry \citep[e.g.,][]{Piette_2020}. We adopt $P$--$T$ profiles consisting of an isotherm down to some radiative-convective boundary (determined by a pressure $P_{\rm rc}$), at which point the model switches to follow an adiabatic profile. $P_{\rm rc}$ serves as an additional input to the model. Higher $P_{\rm rc}$ values extend the isothermal region deeper into the envelope, leading to a cooler temperature profile overall. This mechanism serves as a proxy for planetary age, where lower $P_{\rm rc}$ values correspond to younger and hotter planets, a relationship which has been demonstrated for evolutionary models \citep{Rogers2025}. By varying $P_{\rm rc}$, we can draw qualitative conclusions regarding the time evolution of these planets. Although this is a relatively simple prescription for the $P$--$T$ profile, it has been shown to be adequate for interior structure modeling \citep{Nixon_Madhusudhan_2021, Rogers2025}.

\subsection{Model Grid}

\begin{table}[hbt!]
\centering
\begin{tabular}{c|c|c|c|c}
Parameter & Minimum & Maximum & $N_{\rm points}$ & Scale    \\\hline
$M_p \; (M_\oplus)$ & 4 & 20 & 5 & Linear   \\
$T_0\;$(K) & 300 & 1100 & 5 & Linear    \\
$P_{\rm rc}\;$(bar) & 1 & 100 & 5 & Log    \\
$x_{\rm env}\;$(\%) & 0.1 & 50 & 50 & Log     \\ 
$\mu\;$(g/mol) & 2.35 & 18.02 & 50 & Linear    
\end{tabular}
\caption{\label{tab:grid} Model parameter space; shows the range and spacing for each parameter along with the number of points.}
\end{table}

Our model grid explores the effect of varying five input parameters: planet mass ($M_p$),  pressure at the radiative-convective boundary ($P_{\rm rc}$), photospheric temperature ($T_0$), envelope mass fraction ($x_{\rm env}$), and mean molecular weight (MMW). The precise ranges and spacing of our parameters are described below and shown in Table~\ref{tab:grid}. 

After an initial exploration of a coarse model grid, we conclude that the dependence of the envelope-mantle boundary conditions on $x_{\rm env}$ and MMW is stronger compared to the other parameters under consideration. For this reason, our grid includes more samples of these parameters (50 each) than of $M_p$, $T_0$, and $P_{\rm rc}$ (5 each). In total our grid consists of 312,500 interior structure models. 

The minimum $M_p$ of 4 $M_\oplus$ is an approximate lower bound for sub-Neptunes with substantial gaseous envelopes \citep{Parc_2024}
while the maximum planetary mass of 20 $M_\oplus$ is informed by the fact that Neptune itself is only 17 $M_{\oplus}$, alongside constraints from Kepler on the mass of planetary cores \citep{Owen_2013}. The lower limit for $T_0$ of 300 K is informed by the fact that lower temperatures than this can lead to condensation of H$_2$O and possible stratification of the envelope \citep{Benneke_2024}, which we do not consider in this study. The upper limit is comparable to the hottest sub-Neptune whose upper atmosphere has been constrained with JWST spectroscopy, TOI-421~b \citep{Davenport_2025}. We note that very hot sub-Neptunes ($T_{\rm eq} \gg 1000\,$K) are likely to lose their envelopes due to extreme stellar irradiation \citep{Owen_2017}. $P_{\rm rc}$ ranges from 1 to 100 bar on a logarithmic scale, informed by self-consistent atmospheric models of sub-Neptunes \citep[e.g.,][]{Piette_2020}. We vary $x_{\rm env}$ from 0.1\%--50\%, with 0.1\% serving as a minimum for planets likely to retain gaseous envelopes \citep{Owen_2017} and 50\% as a likely maximum limit for ice accretion highlighted in \cite{Lodders_2003} and \cite{Zeng_2019}. The MMW ranges from 2.35 g/mol (pure H/He) to 18.02 g/mol (pure H$_2$O).

We generate a model for each combination of parameters described above; a subset of these models produce radii which are too large to be sub-Neptunes ($R_p > 4R_{\oplus}$). We do not consider these in our results. Each completed model consists of its complete interior structure profile (including the pressure, temperature, mass, radius, and density of each layer), as well as the resulting planet radius.  From these profiles the pressure and temperature at the envelope-mantle boundary is extracted, which is used to find the resultant phase at this location, i.e., the top of the silicate layer.

\section{Results} \label{sec:results}

\subsection{Initial Case Study: GJ~1214~b} \label{subsec:gj1214b}

\begin{figure}
    \centering
    \includegraphics[width=\linewidth]{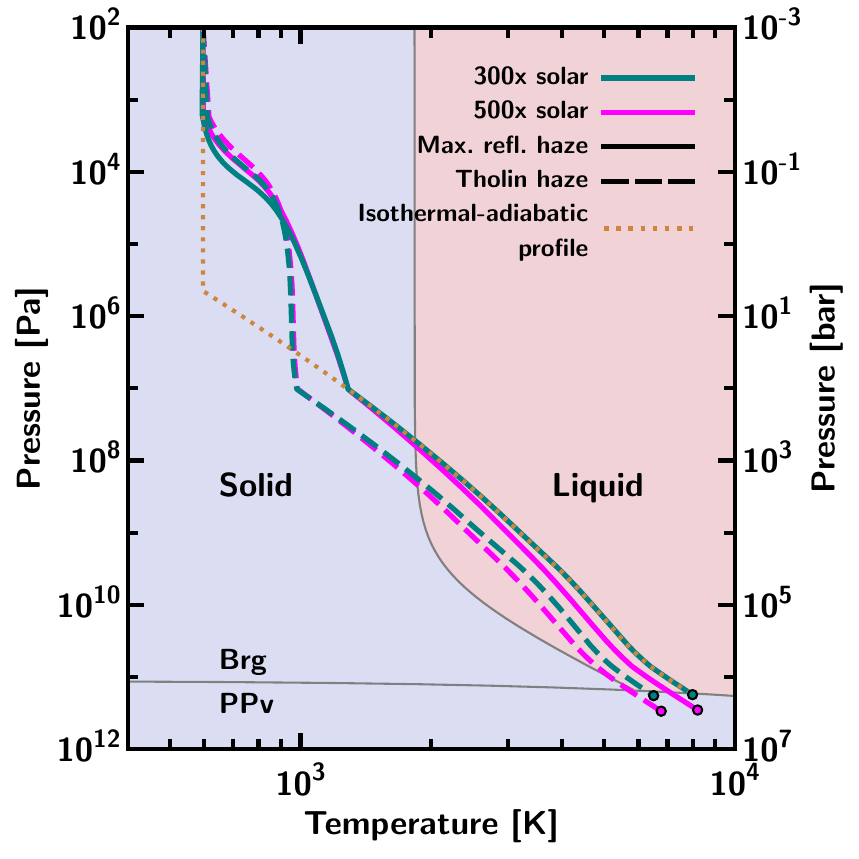}
    \caption{Pressure-temperature profiles of GJ~1214~b with different compositions ($300\times$ and $500\times$ solar metallicity) and haze properties (maximally reflective hazes and tholin hazes), from \cite{Nixon_2024}. The profiles are overlaid on the phase diagram of MgSiO$_3$, showing the solid-liquid and bridgmanite-postperovskite (Brg-PPv) boundaries. Circles indicate the pressure and temperature at the envelope-mantle boundary for each model. In all cases, the boundary occurs in the solid phase. The dotted orange line shows the best-fit isothermal-adiabatic profile (i.e., the parameterization used in this work) to the $300\times$ solar, maximally reflective haze model.}
    \label{fig:GJ 1214b}
\end{figure}

As an initial study, we consider the planet GJ~1214~b \citep{Charbonneau2009}, one of the most extensively studied sub-Neptunes. Observing campaigns from the ground as well as using HST and JWST have determined that this planet hosts a hazy, high-MMW atmosphere \citep{Bean2010,Bean2011,Desert2011,Berta2012,deMooij2012,Teske2013,Kreidberg_2014,Kempton_2023,Gao_2023,Schlawin2024,Ohno2025}. These results, alongside self-consistent radiative transfer calculations to predict the atmospheric structure, were used to inform internal structure models to infer the possible bulk composition of the planet \citep{Nixon_2024}, revealing a substantial gaseous envelope with pressures and temperatures at the envelope-mantle boundary that correspond to a solid (i.e., non-magma) mantle surface. Our investigation of GJ~1214~b serves two main functions: (1) to further explore the extent to which the set of possible internal structures of GJ~1214~b permit a solid envelope-mantle boundary, and (2) to test the validity of the simpler isothermal-adiabatic $P$--$T$ profile assumption that we use in the remainder of this study.

We begin by considering four internal structure models from \citet{Nixon_2024}, covering two envelope compositions (300$\times$ solar and 500$\times$ solar metallicity) and two classes of haze model (maximally reflective hazes and tholin hazes). In each case, the model uses a self-consistent $P$--$T$ profile calculated under the assumption of radiative-convective equilibrium using GENESIS \citep{Gandhi2017,Piette_2020}. We choose appropriate envelope mass fractions from \citet{Nixon_2024} such that the planetary radius is consistent with its measured value from \citet{Mahajan2024}. For each model, we find the pressure and temperature at the boundary between the gaseous envelope and the silicate mantle. The corresponding $P$--$T$ profiles and boundary locations are shown in Figure \ref{fig:GJ 1214b}. We find that in all cases, the thermodynamic conditions at the boundary lead to a solid silicate layer, suggesting that GJ~1214~b is unlikely to possess a magma ocean. We note that the 300$\times$ solar metallicity solutions, which have the lowest metallicity consistent with observations, lie very close to the solid-liquid boundary, meaning that small modeling or observational inaccuracies could move the boundary into the liquid phase. Higher metallicities result in boundary pressures and temperatures more firmly in the solid phase. We conclude that, while it is not possible to definitively confirm that GJ~1214~b does not currently possess a magma ocean, a wide range of plausible bulk and atmospheric properties of this planet are consistent with the absence of a present-day magma ocean.

We use this example case to validate our simplified \mbox{$P$--$T$} profile prescription used in the following section. For each of the four self-consistent $P$--$T$ profiles considered, we find a corresponding isothermal-adiabatic profile that closely matches the deep atmospheric temperature. In order to do this, we fix $T_0$ at the planet's equilibrium temperature \citep[596 K,][]{Cloutier2021} and optimize $P_{\rm rc}$ by conducting a bisection search in log-space until $\Delta T/T < 1\%$ for all pressures in the deep atmosphere ($P>100\,$bar). We verify that the planet radius as well as the temperature and pressure at the envelope-mantle boundary are similar to the values found when using the self-consistent $P$--$T$ profile. Having confirmed this, we justify using the simplified isothermal-adiabatic $P$--$T$ profiles for our large model grid. 

\subsection{Parameter Exploration}

\begin{figure*}[p]
\centering
\includegraphics[width=\linewidth]{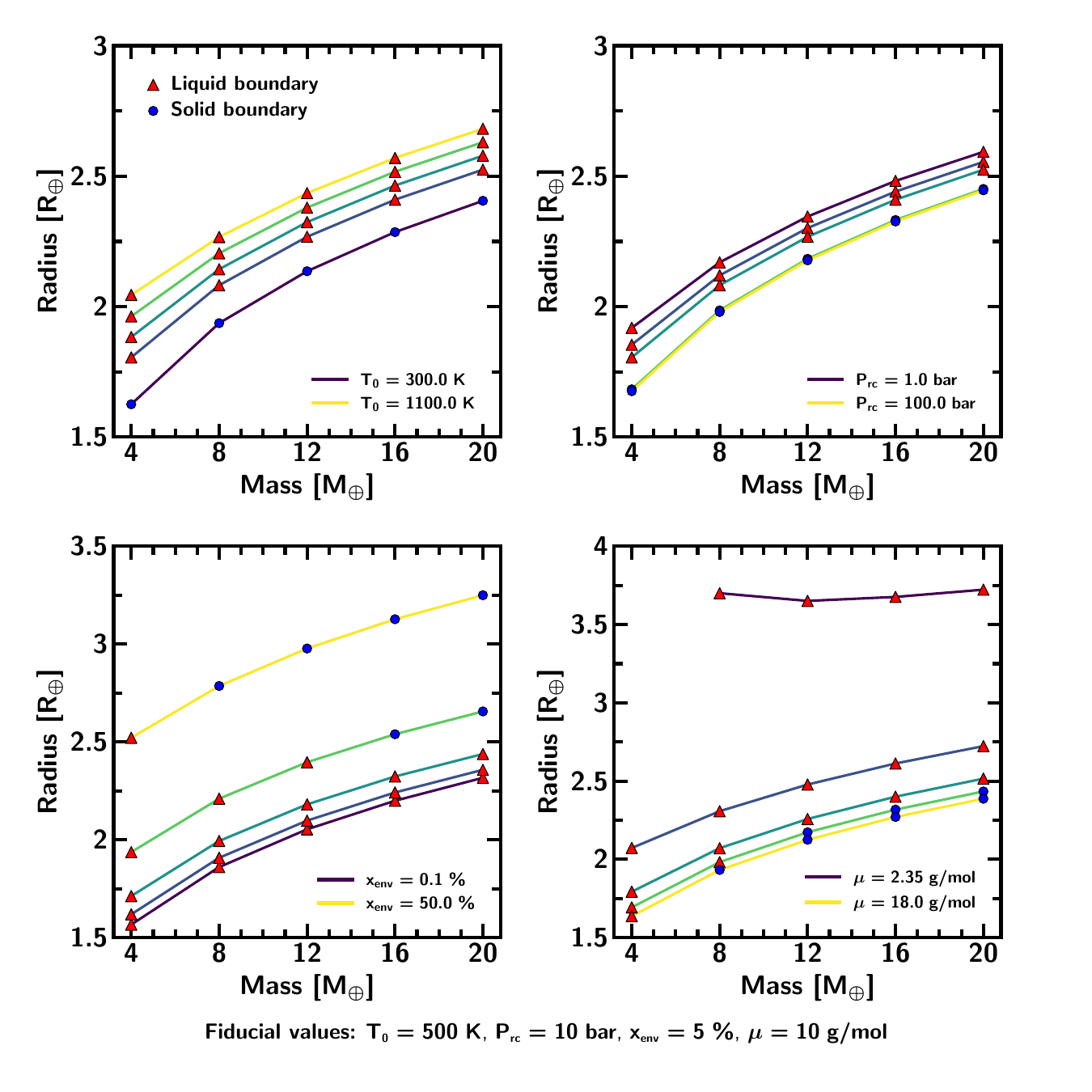}
\caption{Mass-radius curves for a selection of model planets within the model grid. Each panel shows the effect of varying a given parameter on the mass-radius relationship and the presence (or absence) of a magma ocean. The purple line represents the minimum value of each parameter, and the yellow line represents the maximum. Intermediate colors represent evenly-spaced values between the minimum and maximum, using either linear or log-spacing according to Table \ref{tab:grid}. Other parameters are held constant at fixed fiducial values, denoted in the legend. Red triangles indicate model planets with magma oceans, while blue circles indicate a solid silicate layer at the boundary.}
\label{fig:MR_Curves}
\end{figure*}

We now explore the model grid with parameter values shown in Table \ref{tab:grid}, comprising 312,500 models in total. We discount any models that fall under the following criteria: (1) the resulting model radius is greater than 4$R_{\oplus}$, meaning the planet is no longer a sub-Neptune; and (2) the $P$--$T$ profile crosses the H$_2$O liquid-vapor phase boundary. The reason for the second criterion is that our model assumes a mixed H/He/H$_2$O envelope, whereas cooler planets for which this boundary is crossed are more likely to have a stratified envelope due to condensation and cold trapping of H$_2$O \citep{Pierrehumbert2011,Benneke_2024,Gupta2025}. Determining the exact thermal structure (and therefore the thermodynamic conditions at the envelope-mantle boundary) for such planets requires more detailed climate modeling and additional physics beyond what we include in our internal structure model, so we consider these objects to be beyond the scope of our study.

Figure \ref{fig:MR_Curves} shows mass-radius relationships for a selection of model planets in our grid. Although this figure only shows a small subset of models, it is sufficient to provide a qualitative indication of how different parameters can affect the likelihood that a planet could host a magma ocean. We can see that magma oceans appear to persist for planets with high surface temperatures, low pressures at the radiative-convective boundary, low envelope mass fractions, and low MMW. A notable fraction of models which do not meet these criteria have solid silicate layers, meaning no magma ocean. We also see that lower mass planets appear more likely to retain a magma ocean. 

In contrast, planets with both large and small radii can have a solid silicate layer. This follows from the dependence of the radius on other parameters. For example, planets with a high envelope mass fraction will have larger radii, but this leads to a higher probability of a solid surface. Conversely, planets with a higher atmospheric MMW will have smaller radii but are also more likely to host a solid silicate layer. This result highlights that planets across the radius range for sub-Neptunes may not possess magma oceans. Overall, we find that approximately 1/3 of all models in our grid resulted in conditions at the envelope-mantle boundary corresponding to a solid silicate surface with no magma ocean, highlighting that a substantial portion of parameter space exists where magma oceans are not expected (though we note that this fraction depends on the specific grid points that are chosen).

\begin{figure*}[p]
\centering
\includegraphics[width=1\linewidth]{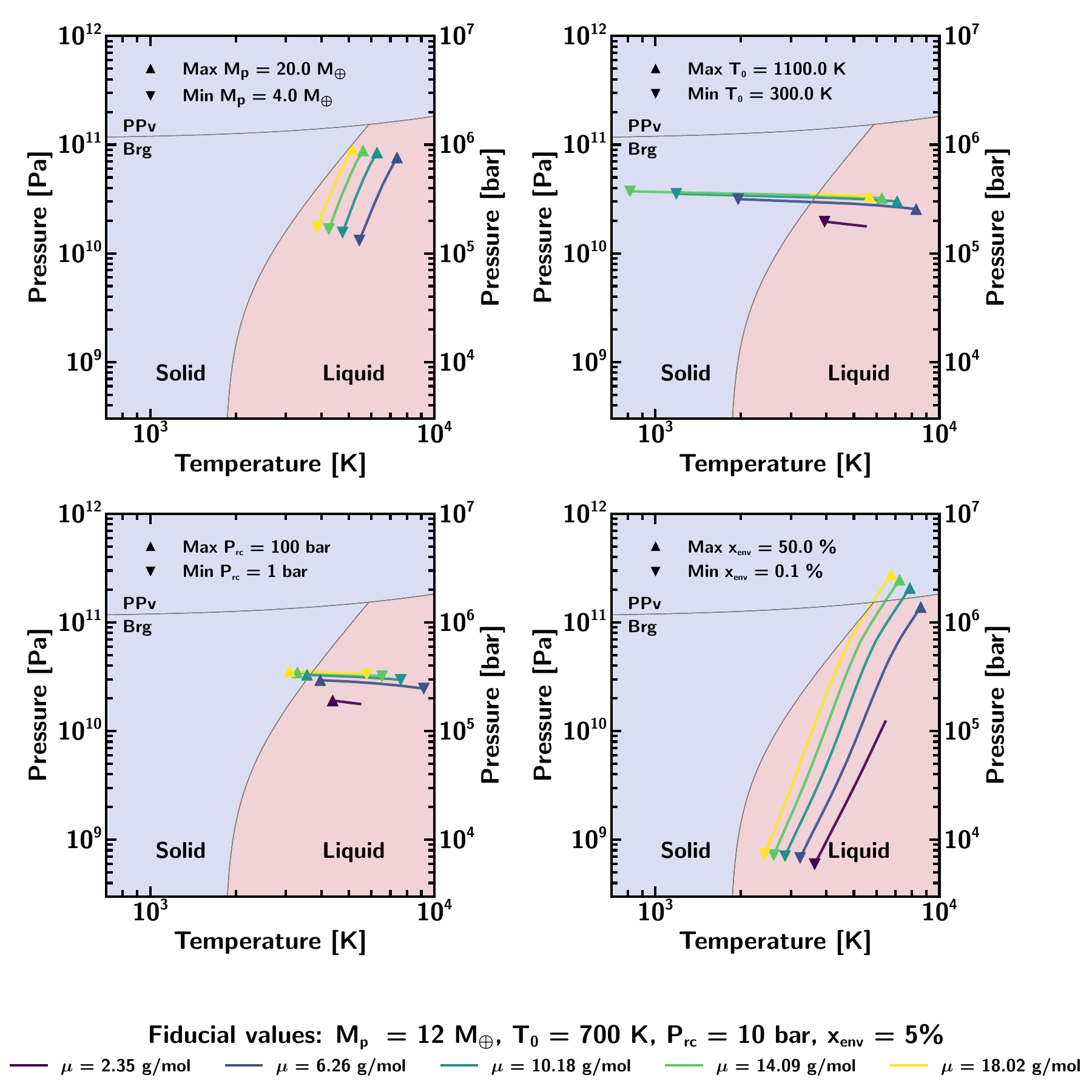}
\caption{Location of the envelope-mantle boundary in pressure-temperature space as a function of parameters within the model grid. Each curve represents the locus of boundary locations across a different parameter range, with an upwards facing triangle representing the maximum value and a downward triangle the minimum. In some cases, the model radius exceeds 4 $R_{\oplus}$ before the maximum (or minimum) is reached when $\mu=2.35\,$g/mol, meaning the following models are not shown at that MMW: $T_0>500\,$K, $P_{\rm rc}<31.6\,$bar and $x_{\rm env}>3.5\%$. Additionally, we do not include the model with $\mu=18\,$g/mol and $T_0=300\,$K as this is impacted by water condensation. The phase diagram of MgSiO$_3$ is shown to indicate the phase at the envelope-mantle boundary for each model in the grid, including the solid-liquid and bridgmanite-postperovskite (Brg-PPv) boundaries. Different line colors represent the MMW of the model. Fiducial values for parameters which are not varied in a given plot are shown below the panels.}
\label{fig:IP}
\end{figure*}

Figure \ref{fig:IP} explores how individual input parameters affect the pressure and temperature at the envelope-mantle boundary in more detail. We show how the location of the boundary moves in $P$--$T$ space as a function of each parameter. We find that each parameter produces its own unique track in this phase space that explains the transitions between magma oceans and solid surfaces seen in Figure \ref{fig:MR_Curves}:

\begin{itemize}
    \item Increasing the planet mass ($M_p$) causes both the pressure and temperature at the envelope-mantle boundary to increase. Although all of the models shown in the upper left panel of Figure \ref{fig:IP} fall in the liquid phase (i.e., the planet has a magma ocean), the increasing boundary pressure with mass can lead to a phase transition, where the pressure becomes sufficiently high that a magma ocean is suppressed.
    \item Increasing the surface temperature ($T_0$) leads to higher temperatures and slightly lower pressures at the envelope-mantle boundary. This means that planets with higher surface temperatures are more likely to host a magma ocean.
    \item Increasing the pressure at the radiative-convective boundary ($P_{\rm rc}$) leads to lower temperatures at the envelope-mantle boundary. Allowing a deeper isotherm leads to a cooler $P$--$T$ profile, meaning that planets with a high $P_{\rm rc}$ are more likely to have a solid silicate surface. Since $P_{\rm rc}$ serves as a proxy for age, this suggests that planets may lose their magma ocean as they evolve (see Section \ref{sec:discussion}).
    \item Increasing the envelope mass fraction ($x_{\rm env}$) leads to higher pressures and temperatures at the envelope-mantle boundary. The impact on pressure is particularly strong, meaning that for high $x_{\rm env}$, the boundary pressure can easily become high enough to prevent the existence of a magma ocean.
    \item Increasing the MMW ($\mu$) leads to lower temperatures and higher pressures at the envelope-mantle boundary. A change in MMW affects the adiabatic gradient of the atmospheric $P$--$T$ profile, with higher-MMW envelopes on a shallower temperature gradient leading to cooler boundary temperatures. The overall result is that high-MMW envelopes are more likely to host a solid silicate surface with no magma ocean.
\end{itemize}

Another takeaway from Figure \ref{fig:IP} is that, due to the nature of the silicate phase diagram, there are two possible ways to suppress a magma ocean: either the boundary can be too cold, leading to the freezing of magma into bridgmanite, or it can be too high-pressure, leading to a phase transition between liquid silicate and solid postperovskite. Many of the model parameters which affect the existence of a magma ocean do so primarily in one of these two ways: for example, $T_0$ and $P_{\rm rc}$ primarily control the temperature at the boundary, whereas $M_p$ and $x_{\rm env}$ strongly affect the pressure. This provides insight into how and why certain sub-Neptunes may not possess magma oceans.

\begin{figure*}[p]
\centering
\includegraphics[width=1\linewidth]
{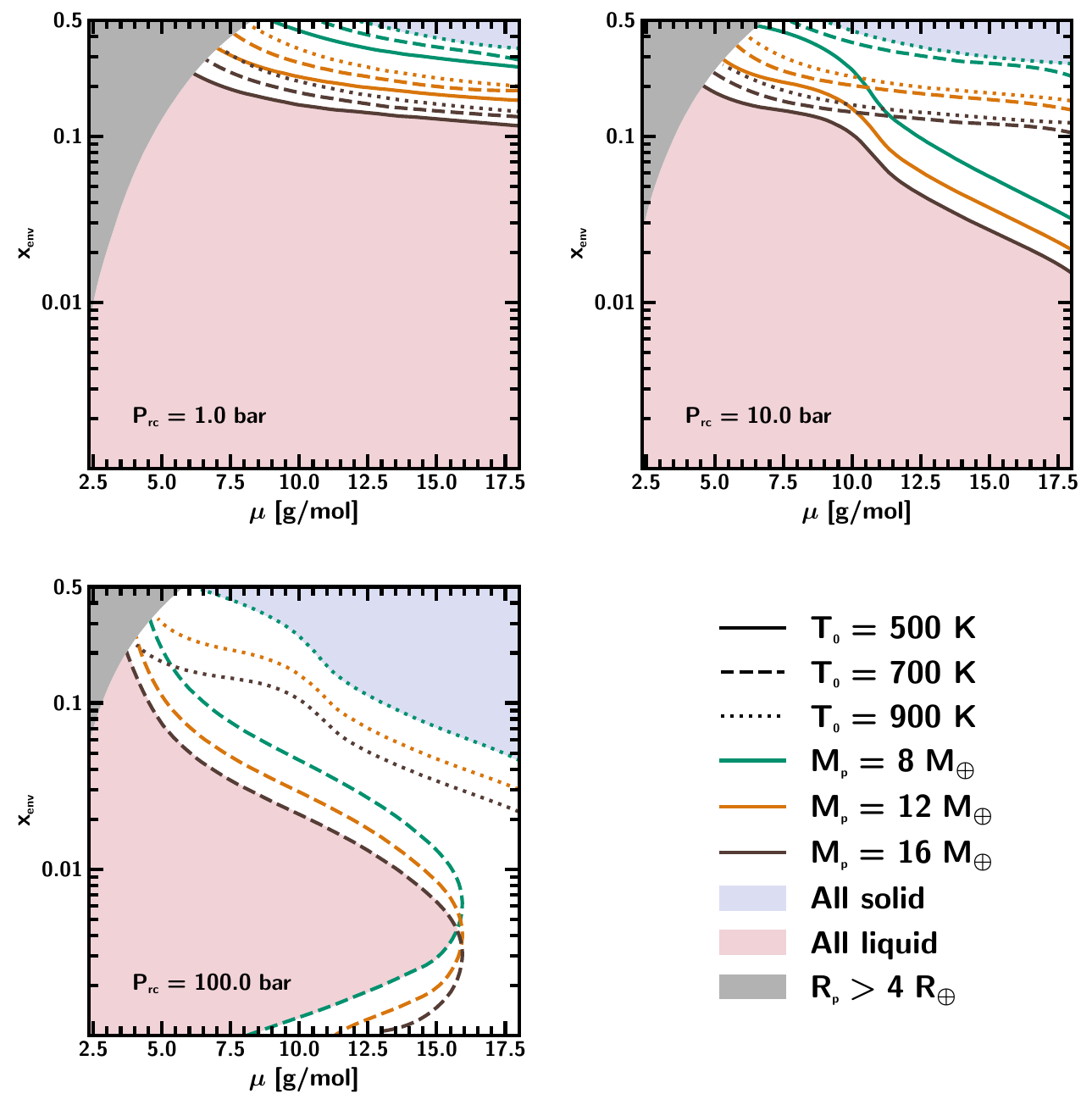}
\caption{Boundaries for the presence or absence of a magma ocean in $x_{\rm env}$--MMW space. Each line represents the boundary with a particular $M_p$ (represented by color) and $T_0$ (represented by line style). Different panels show different values of $P_{\rm rc}$. Models that lie above and to the right of the colored lines lack magma oceans, whereas those that lie to the left and below the lines do possess magma oceans. Blue shading represents regions of the parameter space for which all models feature a solid silicate surface, and red shading represents regions where all models feature a magma ocean (liquid silicate surface). Grey shaded regions represent models whose radius exceeds 4 $R_{\oplus}$. We do not include models with $T_0 = 500\,$K, $P_{\rm rc}=100\,$bar since they are impacted by water condensation.}
\label{fig:MMW_Menv}
\end{figure*}

Given the strong effects of both MMW and $x_{\rm env}$ on the envelope-mantle boundary phase, we explore this particular space in detail. We show the boundary line between the solid and liquid regions of MMW-$x_{\rm env}$ space over a wide parameter range (Figure~\ref{fig:MMW_Menv}). The gray regions on the upper left denote where the output radius exceeds 4 $R_{\oplus}$. High $x_{\rm env}$ and low MMW leads to a larger radius, meaning these planets are no longer considered sub-Neptunes. 

In line with expectations from Figure \ref{fig:IP}, planets with high MMW and $x_{\rm env}$ are the most likely to have solid surfaces. Sub-Neptunes with high $x_{\rm env}$ often lack magma oceans, although the exact cutoff depends on other parameters such as $T_0$. In general, sub-Neptunes with low-MMW envelopes ($\mu \lesssim 5\,$g/mol) are likely to host magma oceans. Planets with low-MMW envelopes and high $x_{\rm env}$ fall outside of the sub-Neptune regime. Higher-MMW envelopes may or may not have magma oceans, depending on $x_{\rm env}$.

Even for planets with $P_{\rm rc}$, $M_p$, and $T_0$ which maximize the likelihood of magma oceans, some combinations of $x_{\rm env}$ and MMW may lead to a solid surface. For example, a planet with $T_0 = 900\,$K, $M_p = 8 M_{\oplus}$, and $P_{\rm rc} = 1\,$bar would not have a magma ocean for $x_{\rm env} > 40$\% and $\mu > 15\,$g/mol. This highlights that even young (low $P_{\rm rc}$), hot (high $T_0$) planets could lack magma oceans if their envelopes are heavy enough. This has implications for the evolution of sub-Neptunes; some of them could spend very little time with a magma ocean, which could dramatically impact atmospheric and structural development \citep[e.g.][]{Chachan_2018, Seo_2024, Rigby_2024}. 

The boundary lines in Figure~\ref{fig:MMW_Menv} between a solid and liquid surface are typically monotonic, with $x_{\rm env}$ decreasing as MMW increases. A notable exception to this occurs at $P_{\rm rc} = 100\,$bar and $T_0 = 700\,$K, where we see an S-shaped turnover in the boundary curve. This means that lower MMW, low $x_{\rm env}$ model planets in this regime also lack magma oceans.  This arises because the temperature profiles for these models (i.e., high $P_{\rm rc}$ and low $T_0$) are simply too cold at their base to sustain a magma ocean, resulting in a solid bridgmanite surface.  This contrasts from the other models shown in Figure~\ref{fig:MMW_Menv}, where the magma ocean is typically suppressed due to very high pressures at the envelope-mantle boundary and a corresponding phase transition to postperovskite. The former behavior only occurs for planets with very thin envelopes, often resulting in planetary radii that are so small that they no longer correspond to sub-Neptunes at all but rather super-Earths ($R_p<1.75\,R_{\oplus}$).  We can conclude that this low-temperature and lower-pressure mantle solidification mode may be unlikely to occur within the sub-Neptune regime of exoplanets.  We note that in Figure~\ref{fig:MMW_Menv} we omit the boundary for $P_{\rm rc} = 100\,$bar, $T_0 = 500\,$K, since this $P$—$T$ profile crosses over the liquid-vapor boundary for H$_2$O and is thus impacted by condensation.

\subsection{Implications for JWST Sub-Neptune Observations}

Our findings indicate that, for a given bulk composition and $P$--$T$ profile, we expect to see a transition between planets hosting or lacking magma oceans which depends on MMW, with lower-MMW atmospheres resulting in magma oceans and higher-MMW atmospheres hosting a solid silicate surface. We use this result to explore how atmospheric MMW could indicate the presence or absence of a magma ocean for a selection of sub-Neptunes with existing or upcoming JWST observations.

Our sample of planets (shown in Table \ref{tab:planets}) includes several sub-Neptunes with published JWST spectra: TOI-421~b \citep{Davenport_2025}; TOI-270~d \citep{Benneke_2024,Holmberg_2024}; GJ~1214~b \citep{Kempton_2023,Gao_2023,Schlawin2024,Ohno2025}; GJ~9827~d \citep{Piaulet-Ghorayeb_2024}; K2-18~b \citep{Madhusudhan_2023,Madhusudhan2025,Hu2025}; and TOI-836~c \citep{Wallack2024}. We include four additional JWST targets whose spectra have not yet been published to fill out our coverage of mass and temperature space: TOI-824~b, TOI-1231~b, HD~207496~b and HD~86226~c.

For each target, we generate interior structure models at its measured mass with MMW values ranging from 2.35--18.02. In each case, we conduct a bisection search to find $x_{\rm env}$ such that the output radius is consistent with the measured radius of the planet. We then identify the conditions at the envelope-mantle boundary in each model to determine whether a magma ocean is present or absent. We repeat this exercise with three $P$--$T$ profiles per planet, each with $T_0=T_{\rm eq}$ and a different value of $P_{\rm rc}$ within the range of our large model grid: 1 bar, 10 bar and 100 bar. This allows us to account for uncertainty in the age and intrinsic temperature of each planet, since $P_{\rm rc}$ acts as a proxy for these quantities. In each case, we search for a transition between the presence or absence of a magma ocean as a function of MMW.

Our results are presented in Table \ref{tab:planets}. In almost all cases, we find a transition between a magma ocean and solid surface at some MMW between a pure H/He and pure H$_2$O atmosphere. This suggests that measuring atmospheric MMW is critical to determining whether a given planet hosts a magma ocean. A small number of cases yield solutions which all possess magma oceans (GJ~9827~d, HD~207496~b and HD~86226~c at $P_{\rm rc}=1\,$bar), and one case shows only solid surfaces (K2-18~b at $P_{\rm rc}=100\,$bar). We note that for other values of $P_{\rm rc}$, these planets do exhibit a transition from magma ocean to solid surface, indicating that for all planets in the sample, the presence of a magma ocean depends on atmospheric properties.

With one exception, we find that lower-MMW atmospheres (i.e., MMW is below the transitional value) possess magma oceans, while atmospheres with a MMW above the transitional value host solid surfaces. This result is in agreement with the findings from our large model grid, shown in e.g. Figure \ref{fig:MMW_Menv}. The exception to this trend is HD~86226~c, for which lower-MMW atmospheres do not have a magma ocean, and higher-MMW atmospheres do. This is because HD~86226~c is the hottest planet in our sample while also having a relatively high density, meaning that at low MMW, the envelope fraction required to explain its mass and radius becomes very small. Since the atmosphere is so thin in these cases, the pressure and temperature at the boundary remain low enough that the silicate layer remains in the solid phase. A similar effect is seen in the lower left panel of Figure \ref{fig:MMW_Menv}, where the transitional MMW for a magma ocean begins to decrease as $x_{\rm env}$ decreases. Recent HST observations of HD~86226~c hint at a high-metallicity atmosphere \citep{Kahle2025}, but further observations using JWST will be required to break the degeneracy with high-altitude clouds.

For planets with measured MMW values from JWST observations, our results help to inform the likelihood of a present-day magma ocean. The transitional MMW for TOI-421~b is much higher than the upper limit of $\sim 2.7\,$g/mol found by \citet{Davenport_2025} for all values of $P_{\rm rc}$, indicating that this planet should still host a magma ocean. For TOI-270~d \citep[$\mu=5.47^{+1.25}_{-1.4}\,$g/mol,][]{Benneke_2024}, GJ~1214~b \citep[$\mu \gtrsim 9.8\,$g/mol,][]{Kempton_2023,Gao_2023}, GJ~9827~d \citep[$\mu=18.02^{+0.20}_{-0.61}\,$g/mol,][]{Piaulet-Ghorayeb_2024} and TOI-836~c \citep[$\mu \gtrsim 6\,$g/mol,][]{Wallack2024}, our transitional MMW values span the reported MMW measurements. This suggests that further observational refinement of the MMW is needed to determine the likelihood of a magma ocean. Alternatively, detailed evolutionary modeling could help to refine the range of possible $P_{\rm rc}$ values (see Section \ref{subsec:evolution}). We note that molecular abundances in the atmosphere of TOI-270~d have been found to be consistent with the result of magma-atmosphere interactions \citep{Nixon2025}. A MMW value has not been reported for K2-18~b. The interpretation of spectroscopic observations of this planet has been the source of considerable debate \citep[e.g.,][]{Schmidt2025,Luque2025,Welbanks2025}. Furthermore, due to the possibility of water condensation and cold trapping, the measured composition of the upper atmosphere of this planet may not be representative of the entire gaseous envelope. In this case, other atmospheric indicators may be used to determine the role of magma-atmosphere interactions in shaping this planet \citep{Rigby_2024,Shorttle_2024}. Overall, given the significant number of sub-Neptunes observed to have high-MMW envelopes, we highlight that present-day magma oceans on sub-Neptunes may not be ubiquitous.

\begin{table*}
\centering
\begin{tabular}{l|c|c|c|c|c|c|l}
\hline
Planet & $M_p \; (M_\oplus)$ & $R_p \; (R_{\oplus})$ & $T_{\rm{eq}} \; (\mathrm{K})$ & $\mu_{\rm 1 \, bar}$ & $\mu_{\rm 10 \, bar}$ & $\mu_{\rm 100 \, bar}$ & Reference \\
\hline
GJ 9827 d   & $3.02^{+0.58}_{-0.57}$ & $1.89^{+0.16}_{-0.14}$ & 620  & (All liquid) & 17.04 & 7.89 & \citet{Passegger2024} \\
TOI-270 d   & $4.78 \pm 0.43$ & $2.19 \pm 0.07$ & 365 & 5.93  & 4.63  & 2.67 & \citet{VanEylen2021} \\
HD 207496 b & $6.10 \pm 1.60$ & $2.25^{+0.12}_{-0.10}$ & 743  & (All liquid) & 16.39 & 8.55 & \citet{Barros2023} \\
TOI-421 b   & $6.70 \pm 0.60$ & $2.64 \pm 0.08$ & 981  & 15.73 & 13.12 & 10.51 & \citet{Krenn2024} \\
HD 86226 c  & $7.25^{+1.19}_{-1.12}$ & $2.16 \pm 0.08$ & 1311 & (All liquid) & 2.67* & 3.32* & \citet{Teske2020} \\
GJ 1214 b   & $8.41^{+0.36}_{-0.35}$ & $2.73 \pm 0.03$ & 596  & 11.16 & 9.20  & 3.32 & \citet{Mahajan2024} \\
K2-18 b     & $8.63 \pm 1.35$ & $2.61 \pm 0.09$ & 284  & 4.63  & 3.32  & (All solid) & \citet{Benneke2019} \\
TOI-836 c   & $9.60^{+2.70}_{-2.50}$ & $2.59 \pm 0.09$ & 665  & 13.12 & 10.51 & 5.28 & \citet{Hawthorn2023} \\
TOI-1231 b  & $15.40 \pm 3.30$ & $3.65^{+0.16}_{-0.15}$ & 330  & 3.97  & 2.67  & 2.67 & \citet{Burt2021} \\
TOI-824 b   & $18.47^{+1.84}_{-1.88}$ & $2.93^{+0.20}_{-0.19}$ & 1253 & 13.12 & 9.85  & 7.89 & \citet{Burt2020} \\
\hline
\end{tabular}
\caption{Selection of 10 sub-Neptunes with current or upcoming JWST observations, including transitional values of MMW, above which the planet no longer hosts a magma ocean for a given $P$--$T$ profile. Three values of $P_{\rm rc}$ are considered for each planet, with $\mu_{X\, \rm{bar}}$ representing the transitional MMW for $P_{\rm rc}= X\,$bar. Models that remain liquid or solid across all conditions are indicated in parentheses. For HD~86226~c, MMWs above the transitional value result in a magma ocean (these values are marked by an asterisk). MMW ($\mu$) is reported in units of g/mol.}
\label{tab:planets}
\end{table*}

\section{Discussion and Conclusions} \label{sec:discussion}

\subsection{Connection to planetary evolution} \label{subsec:evolution}

As mentioned in Section \ref{sec:methods}, the $P_{\rm rc}$ parameter serves as a proxy for planetary age, with older planets having a deeper radiative-convective boundary (and therefore cooling over time). Our research suggests that some sub-Neptunes could therefore lose their magma oceans with time. This is analogous to the crystallization of the Earth's magma ocean, which has been extensively studied \citep[e.g.,][]{Abe1993,Boukare2015,Ballmer2017,Korenaga2023}. Additional research has investigated the process of magma ocean crystallization for terrestrial exoplanets \citep[e.g.,][]{Schaefer2018,Boukare2025}. Given our findings, we suggest that further study into magma ocean crystallization for larger sub-Neptunes, where the conditions at the boundary are likely to be different (i.e., at higher pressures) to terrestrial exoplanets, is crucial to further our understanding of sub-Neptune evolution.

While $P_{\rm rc}$ serves as a useful approximate indicator of a planet's age, formally connecting $P_{\rm rc}$ and age is non-trivial, and requires further assumptions about the planet's formation conditions, the luminosity of the host star, and the cooling timescale of the planet \citep{Rogers2025}. Furthermore, the magma ocean crystallization process itself is highly complex \citep[e.g.,][]{Schaefer2018} and would require much more sophisticated models than those used in this work. For these reasons, we chose to retain $P_{\rm rc}$ as a free parameter in the model, rather than tying our model to any specific assumptions about these factors. A future study could incorporate planetary evolution to determine the length of time for which a planet with certain properties would be expected to retain a magma ocean.

\subsection{Model limitations and caveats}

Understanding the interiors of large exoplanets requires improved data on material behavior at extreme pressures and temperatures. For instance, the EOS of water (H$_2$O) at high $P$ and $T$ remains poorly constrained by experiments \citep{Journaux_2020}. There is also some uncertainty in the silicate EOS and phase boundary locations, although the effect on mass-radius relations has been shown to be small for planets $>2\,M_{\oplus}$ \citep{Huang_2022}. For this initial parameter exploration, we have not considered the possibility of partial melting \citep{Fiquet2010}, instead assuming complete melting at the solid-liquid silicate phase boundary, an assumption which could be revisited in future work. Furthermore, while miscibility studies suggest that hydrogen/helium (H/He) and H$_2$O mixtures approximate ideal mixing under certain conditions \citep{Soubiran_2015}, there are also some regions of parameter space where demixing could be expected \citep{Gupta2025,Howard2025}. Although the majority of models in our grid fall into the regime where a mixed envelope is expected, further refining of this boundary would allow for a better understanding of the structure of cooler sub-Neptunes. Further work should also explore the possibility of H$_2$O and H$_2$ mixing with the core and mantle, which has been proposed for planets in the sub-Neptune mass range \citep{Luo2024, Young25}.

Our models follow the assumption that the planet's envelope consists entirely of H, He and H$_2$O. Although other chemical species, such as CH$_4$ and CO$_2$, have been identified in sub-Neptune upper atmospheres \citep[e.g.,][]{Benneke_2024}, we do not possess high-pressure EOS data for these species and therefore they cannot be included in our models at present. Theoretical and experimental work to determine the high-pressure EOS of additional carbon-, nitrogen-, and sulfur-bearing chemical species would be invaluable to enable more detailed modeling efforts of the internal structures of sub-Neptunes.

Additionally, we do not account for the effects of interactions between the magma ocean and the envelope, which are likely to affect the composition of both components \citep{Schlichting_2022}. This process could, for example, enrich atmospheric MMW \citep{Seo_2024} and limit the total radius of the planet \citep{Kite_2019}. However, given the substantial uncertainties which remain surrounding magma ocean-envelope interactions, due in part to a lack of relevant experimental data \citep{Rigby_2024}, we choose not to model this effect and instead allow the atmospheric composition to vary freely. Further work into refining magma ocean-envelope interaction models would allow for this effect to be incorporated more directly into internal structure models.

\subsection{Conclusions}

Our primary goal in this Letter was to determine whether all sub-Neptunes should be expected to host magma oceans, or whether a subset of such planets should present solid surfaces. Through internal structure modeling, we have found that a significant portion of the population may lack present-day magma oceans.  We provide three main conclusions about the conditions that govern the state of silicates at the envelope-mantle boundary.

\begin{enumerate}
    \item  The envelope mass fraction and MMW most strongly impact whether a planet hosts a magma ocean, due to the impact of these parameters on the pressure at the envelope-mantle boundary.  High pressures bring about a phase transition from liquid magma to solid postperovskite, resulting in mantle solidification.  Planets with a high envelope mass fraction and a high MMW are therefore likely to have a solid silicate surface with no magma ocean.  
    \item Planets with the coldest temperature profiles within our model grid, particularly those with high-MMW envelopes and low envelope mass fractions, provide a second mode for mantle solidification via freezing of liquid magma into solid bridgmanite.  However, to allow for sufficiently cold temperatures at the envelope-mantle boundary, such planets must have very thin envelopes, to the point that many exhibit radii consistent with being super-Earths, rather than sub-Neptunes.  
    \item The pressure at the radiative-convective boundary ($P_{\rm rc}$), which serves as a proxy for planetary age, also plays a key role in establishing whether silicates exist in solid or liquid state at the envelope-mantle boundary.  Lower $P_{\rm rc}$, corresponding to younger planetary age, results in conditions consistent with liquid magma. However, our models show that even relatively young ($P_{\rm rc}$ = 1 bar), hot ($T_0>900\,$K) planets can potentially lack magma oceans if the envelope mass fraction and MMW are sufficiently high.
\end{enumerate}

Our work has significant implications for the population of sub-Neptunes which are currently being observed using JWST. Once the bulk and atmospheric properties of a sub-Neptune have been determined, our results can provide an indication of whether said planet is likely to presently host a magma ocean, as we have demonstrated for several previous and upcoming JWST targets. Given that high-MMW ($\mu \gtrsim 5\,$g/mol) atmospheres have already been observed for several sub-Neptunes, there is a real possibility that a large fraction of this population does not currently possess a magma ocean in contact with its gaseous envelope.

Due to the role of magma oceans in the formation and development of sub-Neptunes, it is vital that future efforts to characterize these planets, particularly those with high-MMW envelopes, account for the possibility that a magma ocean may be absent. Population studies of the upper atmospheres of sub-Neptunes across a wide parameter space could test our theoretical work by determining whether the expected impact of magma oceans on atmospheric properties is more prevalent on those objects which we predict to have present-day magma oceans. Additional observations alongside further modeling developments will enable significant progress in our understanding of the formation, evolution and structure of these mysterious worlds in the coming years.

\begin{acknowledgments}

The authors thank Melissa Hayes-Gehrke and Megan Weiner Mansfield for helpful feedback on an early draft of this manuscript, and the anonymous referee for their useful comments. The authors acknowledge the University of Maryland high-performance computing resources used to conduct the research presented in this paper. This research was supported in part by the AEThER program, funded by the Alfred P.\ Sloan Foundation under grant \#G202114194 and by NASA ADAP 80NSSC19K1014. M.C.N.\ thanks the Heising-Simons Foundation for their funding through the 51 Pegasi b Postdoctoral Fellowship. This research has made use of the NASA Astrophysics Data System and the NASA Exoplanet Archive.

\end{acknowledgments}

\software{\texttt{SMILE} \citep{Nixon_Madhusudhan_2021},
          \textsc{NumPy} \citep{Harris2020},  
          \textsc{SciPy} \citep{Virtanen2020}, 
          \textsc{Matplotlib} \citep{Hunter2007},
          \textsc{Pandas} \citep{pandas}
          }

\bibliography{magma_oceans}{}
\bibliographystyle{aasjournalv7}

\end{document}